\documentstyle[prl,aps,epsfig,multicol]{revtex}

\begin{document}
\title{Electronic and magnetic properties of multishell Co nanowires coated with Cu}
\author{Baolin Wang$^{1,2,3}$$^*$, Xiaoshuang Chen$^2$, Guibin Chen$^2$, Guanghou Wang$^3$, Jijun Zhao$^4$$^{\dagger}$ }
\address{$^1${\it Department of physics, Huaiyin Teachers College, Jiangsu 223001, P.R.China }\\
$^2${\it National Laboratory for Infrared Physics, Shanghai Institute of Technical Physics Chinese Academy of Sciences, Shanghai 200083, P.R. China }\\
$^3${\it National Laboratory of Solid State Microstructures and Department of physics, Nanjing University, Nanjing 210093, P.R. China }\\
$^4${\it Department of Physics and Astronomy, University of North Carolina at Chapel Hill, Chapel Hill, North Carolina 27599, USA}}

\maketitle

\begin{abstract}

The structural, electronic, and magnetic properties of ultrathin Cu-coated Co nanowires have been studied by using empirical genetic algorithm simulations and a tight-binding $spd$ model Hamiltonian in the unrestricted Hartree-Hock approximation. For some specific stoichiometric compositions, Cu atoms occupy the surface, while Co atoms prefer to stay in the interior, forming the perfect coated multishell structures. The outer Cu layers lead to substantial variations of the magnetic moment of interior Co atoms, depending on the structure and thickness of Cu layers. In particular, single Co atom row in the center of nanowire is found to be nonmagnetic when coated with two Cu layers. All the other Co nanowires in the coated Cu shell are still magnetic but the magnetic moments are reduced as compared with Co nanowires without Cu coating. The interaction between Cu and Co atoms induces nonzero magnetic moment for Cu atoms.

\end{abstract}

\begin{multicols}{2}
\section{Introduction}

In recent years, there have been increasing interests in the study of low-dimensional metallic nanostructures because of their unique physical properties and the potential applications in nanoscale materials and devices \cite{device}. The progress in experimental techniques have made it possible to synthesize stable ultrathin metal nanowires with diameter down to several nanometers and of sufficient length \cite{wire1,wire2,wire3,wire4,wire5,wire6,wire7}. For example, Kim's group has successfully fabricated arrays of single-crystalline silver nanowires with 0.4 nm width and  ${\mu}$m-scale length inside the pores of organic template \cite{wire7}. Takayanagi {\em et al.} has observed novel helical multi-shell structures in the suspended ultrathin gold and platinum nanowires \cite{wire2}.

Theoretically, the ultrathin metal wires have been investigated by different groups \cite{tosatti1,tosatti2,tosatti3,tosatti4,bilalbegovic1,bilalbegovic2,bilalbegovic3,wang1,wang2,wang3,wang4,wang5,wang6,kang1,kang2,kang3,auwire,sen,opitz}. Most of those previous studies were focused on their structures, melting and dynamical behavior \cite{tosatti1,tosatti2,bilalbegovic1,bilalbegovic2,bilalbegovic3,wang1,wang2,wang3,wang4,wang5,wang6,kang1,kang2,kang3,sen}, while there are only few {\em ab initio} \cite{tosatti3,tosatti4,wang1,wang2,sen,opitz} or semiempircal \cite{wang5,auwire} calculations on the electronic properties of metal nanowires. So far, our knowledge on the electronic and magnetic properties of metal nanowires are still quite limited. In particular, there is no theoretical exploration on the electronic and magnetic properties of ultrathin bimetallic nanowires, whereas the bimetallic nanoclusters have been studies in previous works \cite{bi1,bi2,bi3,bi4,bi5,bi6,alloy1,alloy2,alloy3,jl1,jl2}.

In this paper, we exploit the structural and magnetic properties of Co nanowires fully coated with Cu atomic shells. The main reason for choosing Cu-Co system is that the physical properties of bulk Co and Cu are very different while their atomic radiuses are comparable. We may get a clear dependence of the various properties of the coated nanowires on their size, structure, and thickness of Cu and Co layers. Moreover, Co and Cu alloys are non-miscible. The detailed study of Cu-Co alloy nanowires might provide a qualitative analysis from the mecoscopical points. Previously, the magnetic properties of both free Co clusters \cite{coexp,coexp2,lizq} and the Co clusters embedded in Cu matrix have been studied \cite{cocuexp,cocu1,cocu2,cocu3,cocu4,cocu5}.

\section{Structures of Cu-Co alloy nanowire}

The interatomic interactions are modeled by a Gupta-like many-body potential \cite{gupta}, which was widely used to study the structures and properties of small transition metal clusters \cite{cluster1,cluster2,cluster3} and alloy clusters \cite{alloy1,alloy2,alloy3,jl1,jl2}. The potential parameters for Co-Co and Cu-Cu interaction are taken according to those fitted for bulk Co and Cu solids in \cite{pot}. The parameters for inhomogeneous Cu-Co interaction are derived from the average of the Cu-Cu and Co-Co parameters. In our previous works \cite{jl1,jl2}, the validity of this parameterization scheme was examined by comparing with {\em ab initio} calculations on the small clusters, i.e., homogeneous and inhomogeneous dimers and trimers, Co$_{13}$ and Cu$_{13}$. A direct comparison of the {\em ab initio} result with the empirical one on the structural information, such as equilibrium bond length and bond angle of the small clusters, can be found in Ref.\cite{jl1,jl2}. This parameterization has successfully described the structural electronic, magnetic and thermal properties of Co$_{18-m}$Cu$_m$($0\leq m\leq 18)$ clusters\cite{jl1,jl2}.

In this work, the lowest energy structures of the Cu-coated Co nanowires in several specific stoichiometric compositions are determined using a genetic algorithm (GA) \cite{ga0,ga1,ga2,wang1,wang2,wang3} global structural minimization. In breif, a number of random initial configurations of the nanowires within one-dimensional periodic boundary condition are generated in the beginning. Any two candidates in the population are then chosen as parents to generate a child cluster by mating operation \cite{ga0,wang1}. The child nanowire structure is further relaxed using empirical molecular dynamics. Then the optimized structure will be selected to replace its parents if it has lower energy. The detailed description of the genetic algorithm in the structural optimization of low-dimensional nanostructures can be found in our previous publications \cite{ga1,ga2,wang1,wang2,wang3}.

From our simulations, we found that the Cu-Co alloy nanowire tend to form ordered structures. Cu atoms prefer to occupy the surface while Co atoms stay in the interior of wire. The assembling of Cu atoms tends to maximize the number of Cu-Cu and Co-Co bonds. This phenomenon can be attributed to the significant difference in the surface energy and the cohesive energy between Co and Cu. The cohesive energy and surface energy of the bulk Cu solid are 3.544 eV per atom and 1.934 Jm$^{-2}$ respectively. Both are smaller than those of the Co solid, 4.386 eV and 2.709 Jm$^{-2}$. To minimize the total energy, the atom with a smaller surface energy and cohesive energy tends to occupy the surface, while the atom with a higher surface energy and cohesive energy favors to stay in the interior.  Another possible reason is the atomic size effect. In our simulations, the first neighboring distance of Cu is 2.556 {\AA}, larger than that of Co (2.507 {\AA}). Thus, Co atoms are more easily surrounded by Cu atoms. Basing on these effects, we can adjust the composition ratios of Cu and Co to achieve some perfect-coated structures, for which their electronic and magnetic properties are studied.

Multi-shell cylindrical structures are typically found for the Cu-Co alloy nanowires in our GA simulations. These structures are multi-shell central cylinders composed of regular coaxial cylindrical shells. Such structures have been theoretically predicted for Al, Pb, Au, Ti, Zr, Rh, Cu nanowires \cite{tosatti1,wang1,wang2,wang3,wang5,kang1} and experimentally observed in Au and Pt nanowires \cite{wire2}. On the other hand, coaxial nanocable by semiconducting materials were reported by Iijima's group \cite{nanocable}. In this paper, we restrict our discussions on those nanowires belonging to the centered pentagonal and hexagonal multishell growth sequence because of their relatively higher symmetry and thermal stability \cite{wang6}. Our systematical illustration of various multishell growth sequence of the metal nanowires can be found elsewhere \cite{wang2,wang3}.

\end{multicols}
\begin{figure}
\vspace{-0.1in}
\centerline{
\epsfxsize=5.15in \epsfbox{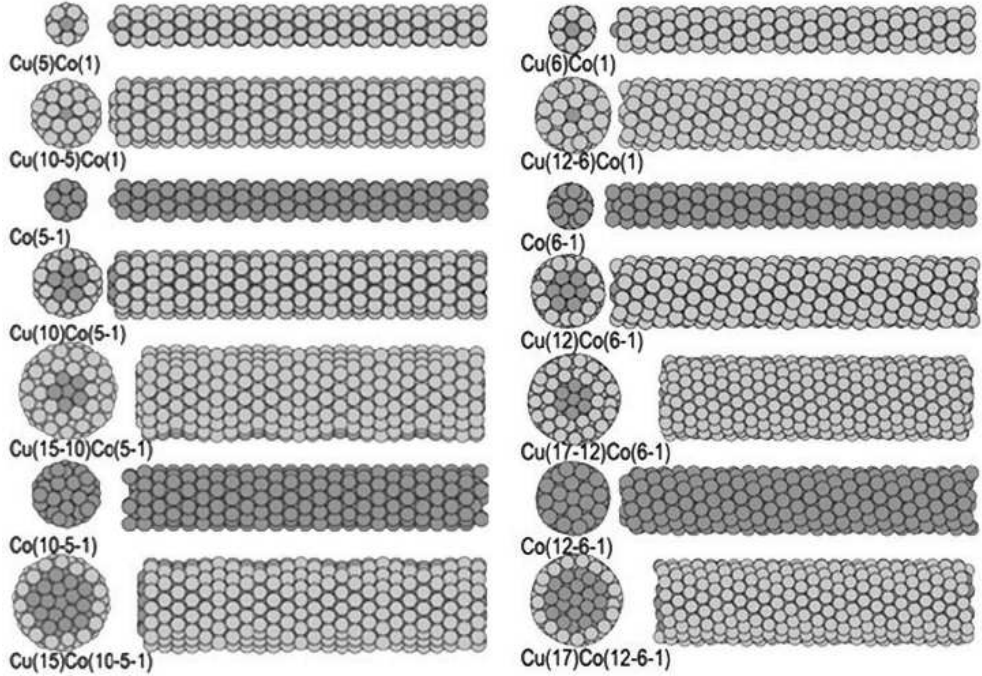}
}
\vspace{0.1in}
\caption{Optimized structures of bare and Cu-coated Co nanowires  (dark balls: Co atoms; bright balls, Cu atoms). Left part is for the centered hexagonal multishell structural pattern and right one for the centered pentagonal one. The indexes of $n$-$n1$-$n2$-$n3$ in parenthesis are used to characterize each structure. See text for detailed descriptions.}
\end{figure}

\begin{multicols}{2}

Fig.1 shows the morphologies of the Cu-Co alloy nanowires with centered pentagonal and hexagonal multishell sequence. Both the bare Co nanowire and the Co wires perfectly coated with Cu are presented. With the notation of $n$-$n1$-$n2$-$n3$ in parenthesis \cite{wire2,wang1,wang2,wang3}, we can describe the nanowires consisting of closed coaxial shells, each of which is composed of $n, n1, n2$, and $n3$ helical atomic strands from outer to inner ($n > n1 > n2 > n3$)\cite{wang2,wang3}. As shown in Fig.1, wire Cu(6)Co(1) is formed by the outer shell containing six Cu helical atomic strands and the inner shell being a single row of Co atoms. Similarly, wires Cu(12-6)Co(1), Co(6-1), Cu(12)Co(6-1), Cu(17-12)Co(6-1), Co(12-6-1), and Cu(17)Co(12-6-1) constitute growth patterns with two-, three-, and four-shells sequence of centered hexagonal packing. In contrast to the helical structures found in the centered hexagonal growth pattern, the Cu(5)Co(1), Cu(10-5)Co(1), Co(5-1), Cu(10)Co(5-1), Cu(15-10)Co(5-1), Co(10-5-1), and Cu(15)Co(10-5-1) nanowires associated with the centered pentagonal multishell sequence are not helical. These structures can be viewed as deformed icosahedral packing. 

\section{Electronic structures and magnetism of Cu-coated Co nanowires}

Based on the optimized structures in Fig.1, we investigate the electronic and magnetic properties by a parameterized $spd$ tight-binding model Hamiltonian within the unrestricted Hartee-Fock approximation. Similar tight-binding methods were successfully applied to transition-metal clusters \cite{jl1,jl2,va,alfe} and nanowires \cite{wang5} in our previous works. The Hamiltonian, written in a local orbital basic set, has the expression:
\begin{equation}
H=\sum\limits_{i,\alpha ,\sigma }\epsilon _{i\alpha \sigma }\stackrel{\wedge 
}{n}_{i\alpha \sigma }+\sum\limits_{\stackrel{i\not{=}j}{\alpha ,\beta
,\sigma }}t_{ij}^{\alpha \beta }\stackrel{\wedge }{c}_{i\alpha \sigma }^{+}%
\stackrel{\wedge }{c}_{i\beta \sigma }
\end{equation}
where $\stackrel{\wedge }{c}_{i\alpha \sigma }^{+}$( $\stackrel{\wedge }{c}_{i\beta \sigma }$) is the creation (annihilation) operator and $\stackrel{\wedge }{n}_{i\alpha \sigma }$is the number operator of an electron.. The $t_{ij}^{\alpha \beta }$is the hopping integral between different sites and orbitals. The orbital state $\alpha $ involved in the calculation includes $s,p_x,p_y,p_z,d_{xy}$,$d_{yz}$,$d_{xz}$,$d_{x^2-y^2}$,$d_{3z^2-r^2}$. The single-site energy $\epsilon _{i\alpha \sigma }$ is given by 
\begin{equation}
\epsilon _{i\alpha \sigma }=\epsilon _d^0+U\Delta n(i)-\frac 12\sigma J\mu
(i)+\sum\limits_{j\not{=}i}\Delta n(j)V_{ij}
\end{equation}
Here $\epsilon _d^0$ refers to the orbital energy levels in the paramagnetic solutions of the bulk. $\Delta n(j)$ denotes the charge change. The Coulomb interaction $V_{ij}$ is described as
\begin{equation}
V_{ij}=\frac U{1+(UR_{ij}/e^2)}
\end{equation}

During our calculations, the orbital energy and the hopping integrals are taken to be the bulk values obtained from Andersen's linear muffin-tin orbital atomic sphere approximation (LMTO-ASA) paramagnetic bands\cite{Andersen}. The hopping integrals are assumed to be spin independent and are averaged for the heteronuclear. Exchanging integrals other than $J_{dd}$ are neglected and $J_{dd}(Co)=0.99$eV\cite{jdd}. The direct integral $U_{dd}(Co)$ is obtained from Ref.\cite{udd} , and $U_{ss}/U_{dd}$ relations are from the atomic tables. We take $U_{dss}=U_{pp}=U_{sp}$ and $U_{sd}=U_{pd}=(U_{ss}+U_{dd})/2$\cite{fab}. For Cu, all the parameters come from Ref.\cite{fab}.

The magnetic moment can be determined by integrating the majority and minority local densities of state(LDOS) up to Fermi energy: 
\begin{equation}
\mu _{i\alpha }=\int_{-\infty }^{\epsilon _F}[\rho _{i\alpha _{\uparrow
}}(\epsilon )-\rho _{i\alpha _{\downarrow }}(\epsilon )]d\epsilon \text{,}
\end{equation}

The LDOS is directly related to the diagonal elements of the local Green function by means of the recursion method\cite{recursion}: 
\begin{equation}
\rho _{i\alpha \sigma }=-\frac 1\pi \mathop{\rm Im} [G_{i\alpha \sigma ,i\alpha \sigma }(\epsilon )]\text{.}
\end{equation}

Our main theoretical results are summarized in Table I, including the average coordination numbers (CN), average bond length $R$, average magnetic moments per Co and Cu atom for bare and Cu-coated Co nanowires. The $n$-$n1$-$n2$-$n3$-$n4$ structural indexes for these wires are given in parenthesis. To simplify the discussion, we will divide the $N$ atoms in the simulation supercell of nanowire into several groups based on the atomic layers, coordination numbers and symmetries of the wire and take their average (see Table II). For example, there are two group of atomic sites (i.e., centered single atom-row and outer shell) in Co(5-1), Cu(5)Co(1), Co(6-1), and Cu(6)Co(1) wires, and four in Cu(10-5)Co(1), Cu(10)Co(5-1), Cu(12-6)Co(1), and Cu(12)Co(6-1).

\begin{table}
Table I. The average coordination numbers (CN), average bond length R({\AA}), calculated average magnetic moments per Co ($\mu _{Co}$) and Cu ($\mu _{Cu}$) atom (in unit of $\mu _b$) for Cu-Co nanowires. The structural index $n$-$n1$-$n2$-$n3$-$n4$ for these wires is given in parenthesis.

\begin{center}
\begin{tabular}{ccccc}
Nanowire & CN & R & $\mu _{Co}$ & $\mu _{Cu}$  \\ \hline
Cu(5)Co(1) & 8.67 & 2.565 & 0.68 & 0.06 \\ 
Cu(10-5)Co(1) & 9.81 & 2.559 & 0.06 & 0.02 \\ 
Co(5-1) & 8.67 & 2.527 & 1.02 & --- \\ 
Cu(10)Co(5-1) & 9.81 & 2.539 & 1.66 & 0.05 \\ 
Cu(15-10)Co(5-1) & 10.31 & 2.529 & 0.14 & -- \\ 
Co(10-5-1) & 9.81 & 2.511 & 0.83 & --- \\ 
Cu(15)Co(10-5-1) & 10.31 & 2.518 & 0.51 & -- \\ 
Cu(6)Co(1) & 8.86 & 2.551 & 1.01 & 0.13 \\ 
Cu(12-6)Co(1) & 9.27 & 2.536 & -0.01 & 0.04  \\ 
Co(6-1) & 8.86 & 2.514 & 2.15 & --- \\ 
Cu(12)Co(6-1) & 9.27 & 2.516 & 1.55 & -- \\ 
Cu(17-12)Co(6-1) & 10.00 & 2.537 & 0.05 & -- \\ 
Co(12-6-1) & 9.27 & 2.492 & 1.63 & ---  \\ 
Cu(17)Co(12-6-1) & 10.00 & 2.519 & 0.45 & -- \\ 
\end{tabular}
\end{center}
\end{table}

First, it is interesting to discuss the magnetic properties of bare Co nanowires with no Cu outer layers. For both hexagonal and pentagonal structural pattern, the magnetic moments of Co nanowires decreases with increasing wire diameters. For example, the average magnetic moments drop from 2.15 $\mu _b$ in Co(6-1) wire to 1.63 $\mu _b$ Co(12-6-1). The trend is similar to the observed magnetic moments of Co clusters \cite{coexp}. It is worthwhile to note that the calculated average magnetic moment of Co(6-1) wire (2.15 $\mu _b$) is close to the magnetic moment of Co$_{13}$ cluster with comparable size. The experimental moment for Co$_{13}$ is 2.08 $\mu _b$ \cite{coexp2}, while first principle calculations on Co$_{13}$ with D$_{3d}$ symmetry give a moment of 2.105 $\mu _b$ per atom \cite{lizq}. However, the magnetic properties for Co nanowires depend not only on the sizes but also on the atomic structures. The magnetic moment found for Co(6-1) wire (2.15 $\mu _b$ per atom) is much higher than the 1.02 $\mu _b$ for Co(5-1) wire, although the sizes of these two wires are comparable. Similar effect can also be seen in the thicker bare Co nanowires, i.e., 2.08 $\mu _b$ for Co(12-6-1) versus 2.08 $\mu _b$ for Co(10-5-1). The larger magnetic moments in the hexagonal nanowires can be attributed to their highly compact helical structures. To achieve a good packing along the wire axis with one-dimensional periodical boundary condition, the match of icosahedral symmetry in pentagonal Co(5-1) or Co(10-5-1) wires will lead to the larger Co-Co interatomic distance as compared with Co(6-1) and Co clusters. The larger bond length would then reduce the magnetic moments \cite{liuf}. 

As shown in Table I, quenching of magnetism in Co nanowires is generally found after Cu layers coating. As an extreme case, single Co atomic row coated by two Cu layers is found to be nearly nonmagnetic (wire Cu(10-5)Co(1) and Cu(17-12)Co(6-1) in Table I). Cu atomic layers seem to have a ''screen'' effect on the magnetic moments of Co atoms and lead to zero magnetic moment in these nanowires. Our current results of Cu-Co nanowires are in quantitatively agreement with previous studies on the Co clusters embedded in Cu matrix. The quenching of ferromagnetism in Co clusters embedded in Cu matrix were reported both experimentally \cite{cocuexp} and theoretically \cite{cocu1,cocu2,cocu3,cocu4,cocu5}. Xiao {\em et al.} even found that single Co atom in Cu matrix is nonmagnetic, similar to our result on the zero magnetic moment for single Co atomic row coated by two Cu layers \cite{cocu1}. 

\end{multicols}

\begin{table}
Table II. Mulliken populations, net charge, and local magnetic moments (in $\mu _b$) for atoms at different shells of nanowires. The $n$-$n1$-$n2$-$n3$-$n4$ and the number for equivalent atomic sites for these wires are given in parenthesis.

\begin{center}
\begin{tabular}{ccccccccc}
Supercell &  &  & Charge &  & Net &  & Moment &  \\ 
&  & 3d & 4s & 4p & charge & 3d & 4s & 4p \\ \hline
Cu(5)Co(1) & core(4 Co) & 8.125 & 0.599 & 0.580 & -0.304 & 0.677 & -0.011 & 
0.017 \\ 
& 2nd(20 Cu) & 9.644 & 0.693 & 0.603 & 0.060 & 0.075 & -0.009 & -0.006 \\ 
Cu(10-5)Co(1) & core( 4Co) & 9.161 & 0.343 & 0.425 & -0.929 & -0.051 & -0.003
& -0.011 \\ 
& 2nd(20 Cu) & 9.531 & 0.508 & 0.929 & 0.032 & -0.039 & 0.000 & 0.021 \\ 
& 3rd(10 Cu) & 9.690 & 0.539 & 0.727 & 0.144 & -0.028 & -0.007 & 0.042 \\ 
& 3rd(20 Cu) & 9.738 & 0.686 & 0.569 & 0.007 & -0.021 & -0.033 & 0.014 \\ 
Cu(10)Co(5-1) & core(4 Co) & 6.121 & 0.609 & 1.280 & 0.990 & 2.534 & -0.028 & 
0.086 \\ 
& 2nd(20 Co) & 7.501 & 0.588 & 0.803 & 0.109 & 1.565 & -0.020 & -0.022 \\ 
& 3rd(10 Cu) & 9.476 & 0.838 & 0.834 & -0.148 & 0.012 & -0.018 & -0.013 \\ 
& 3rd(20 Cu) & 9.551 & 0.859 & 0.752 & -0.162 & 0.027 & 0.016 & 0.023 \\ 
Cu(12-6)Co(1) & core(4 Co) & 9.363 & 0.405 & 0.326 & -1.094 & 0.143 & -0.007
& 0.000 \\ 
& 2nd(20 Cu) & 9.561 & 0.476 & 0.899 & 0.064 & -0.011 & -0.013 & 0.000 \\ 
& 3rd(10 Cu) & 9.620 & 0.610 & 0.744 & 0.026 & -0.020 & -0.017 & 0.046 \\ 
& 3rd(20 Cu) & 9.677 & 0.566 & 0.699 & 0.058 & -0.022 & 0.011 & 0.054 \\ 
Co(12-6-1) & core(4 Co) & 6.646 & 0.649 & 1.462 & 0.243 & 1.261 & -0.054 & 
-0.051 \\ 
& 2nd(28 Co) & 6.966  & 0.573 & 1.348 & 0.113 & 1.924 & -0.062 & -0.044 \\ 
& 3rd(28 Co) & 7.974 & 0.591 & 0.980 & -0.545 & 0.396 & -0.058 & -0.018 \\ 
& 3rd(28 Co) & 6.234 & 0.914 & 1.467 & 0.385 & 2.801 & -0.001 & 0.011 \\ 
Cu(12)Co(6-1) & core(4 Co) & 5.530 & 0.701 & 1.359 & 1.410 & 3.362 & -0.046 & 
-0.090 \\ 
& 2nd(28 Co) & 7.605 & 0.523 & 0.707 & 0.165 & 1.340 & -0.025 & -0.005 \\ 
& 3rd(28 Cu) & 9.468 & 0.826 & 0.914 & -0.208 & 0.024 & -0.025 & 0.024 \\ 
& 3rd(28 Cu) & 9.498 & 0.796 & 0.880 & -0.174 & -0.003 & -0.019 & -0.005 \\ 
Cu(17-12)Co(6-1) & core(4 Co) & 6.327 & 0.534 & 1.066 & 1.073 & 0.331 & -0.002
& 0.042 \\ 
& 2nd(28 Co) & 7.689 & 0.496 & 0.669 & 0.146 & -0.126 & -0.008 & 0.028 \\ 
& 3rd(56 Cu) & 9.548 & 0.617 & 0.938 & -0.103 & -0.009 & -0.002 & 0.010 \\ 
& 4th(81 Cu) & 9.722 & 0.598 & 0.716 & -0.036 & -0.032 & -0.003 & 0.026 \\ 
\end{tabular}
\end{center}
\end{table}
\begin{multicols}{2}

In the case of single Cu layer coating, Co nanowires are still magnetic with reduced moments as compared with the corresponding bare ones (see Table I), with exception at Cu(10)Co(5-1) wire. The average magnetic moment per Co atom in wire Cu(10)Co(5-1) is much larger than one of Co(5-1). The reason might be that Cu layer ''compresses'' the bond length of Co-Co in the Co wire, which lead to the larger charge transfer between Cu and Co atoms. We find the average bond length of the core Co(5-1) wire is 2.516 {\AA}, which is smaller than that 2.527 {\AA} of bare Co(5-1) wire. In most cases, the average magnetic moment of Cu atom is nearly zero, whereas small moments on Cu sites are found in the Cu(6)Co(1), Cu(5)Co(1), Cu(10-5)Co(1), and Cu(10)Co(5-1) wires.

To further explore the interaction between Co core and coated Cu layers, Table II present the Mulliken populations for several bare and Cu-coated Co nanowires. Considerable charge transfer is found in all the nanowire studied. Since the electronegativity for Cu is close to that for Co, it is expected that electrons may transfer either from Co to Cu or from Cu to Co, depending upon the specific structure and environment of the atoms. We find that the charge transfer from Cu atoms to Co atoms occurs for the single atomic Co row with Cu coating, while the charge transfer from Co atoms to Cu atoms is found for the other nanowires. From Table II, electrons are found to transfer from Cu layers to Co row in the Cu(5)Co(1), Cu(10-5)Co(1), and Cu(12-6)Co(1). But for the nanowires Cu(10)Co(5-1), Cu(12)Co(6-1), and Cu(17-12)Co(6-1) there are electrons transferring from Co to Cu. Generally, charge transfer from Cu to Co will depress the magnetic moment of Co atom. On the contrary, when the charge transfers from Co to Cu, an enhanced magnetism of Co will be induced. For example, it is found that the central Co row of nanowires has an enhanced moment when covered with Cu shell, such as Cu(10)Co(5-1) and Cu(12)Co(6-1).

Charge transfer between Co atoms is also found for the bare Co nanowires, for example, Co(12-6-1) in Table II, where electrons transfer to the high-coordinate sites. The transfer is related to a corrugated outmost shell with shorter bond lengths. In addition, there is also considerable charge redistribution between different shells of Co atoms and among different atomic orbitals of the same atom in all nanowires with coating Cu shells. For Co atoms, the charges are transferred from $sp$ orbitals to $d$ orbital, while for Cu atoms, the charges are transferred from $d$ orbital to $sp$ orbitals. 

Fig.2 shows the total density of states (DOS) and the $sp$, $d$ partial DOS of several bare and coated Co nanowires . In general, the DOS near to Fermi level plays a primary role in determining the magnetism of the nanowires. It is clearly seen that the contribution from $d$ electrons is dominant, while the $sp$ components are less important. The contribution of $d$ electrons in Cu(10)Co(5-1) is larger than that in Co(5-1), which leads to the enhanced magnetic moment in Cu(10)Co(5-1) wire. For Cu(10-5)Co(1) and Cu(12-6)Co(1) wires, the contribution of $d$ electrons near to the Fermi level is relatively smaller than that of other cases, which leads to especially low magnetic moments. Moreover, Fig.2 shows that the $sp$-$d$ hybridizations in Cu(10-5)Co(1) and Cu(12-6)Co(1) wires are also less pronounced than other cases, while the hybridizations is much pronounced in Cu(10)Co(5-1) and Cu(12)Co(6-1) wires. This $sp$-$d$ hybridizations also enhances the magnetism of these Cu-Co alloy nanowires. The calculated DOS's for the Cu-Co nanowires show some similarity to those of Cu-Co alloy clusters in Ref. \cite{cocu1}.

\begin{figure}
\vspace{0.75in}
\centerline{
\epsfxsize=4.in \epsfbox{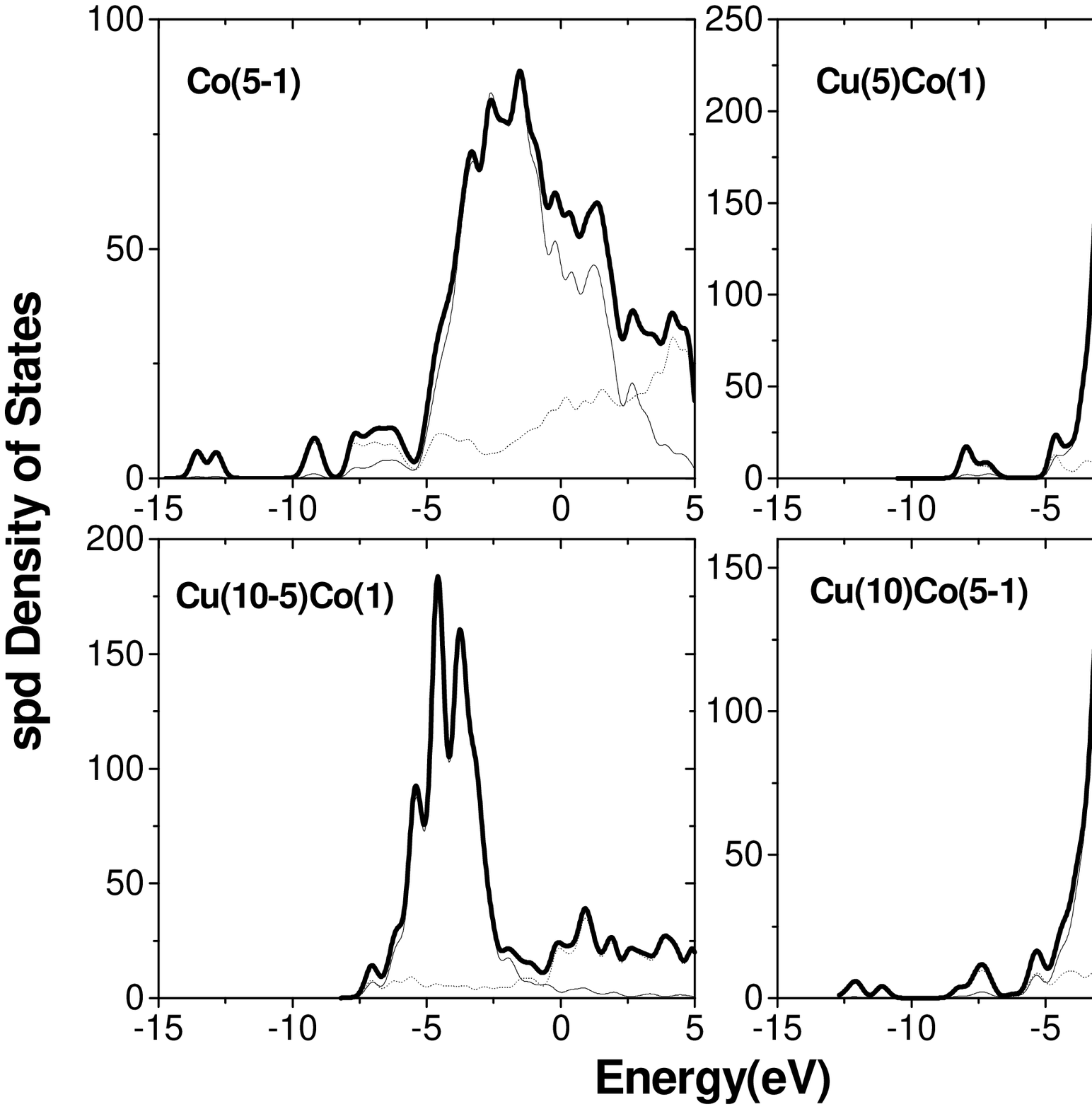}
}
\vspace{0.0in}
\centerline{
\epsfxsize=4.in \epsfbox{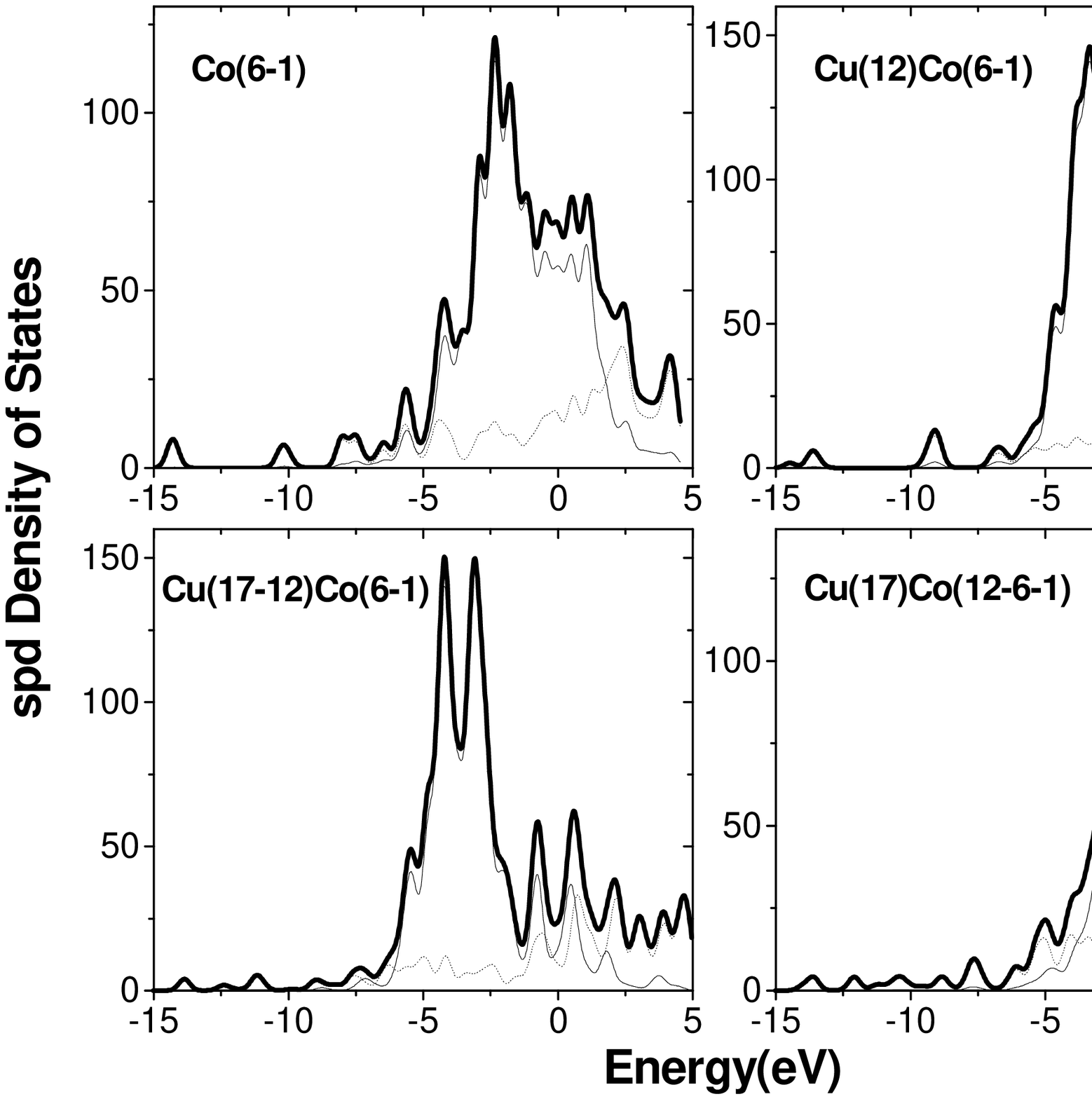}
}
\vspace{-0.7in}
\caption{Total (thick solid line), $sp$ (thin dotted line) and $d$ orbital (thin solid line) density of states for Cu-Co nanowires. Upper: DOS for pentagonal nanowires; down: DOS for hexagonal nanowires. The Fermi level is set as zero. 0.2 eV Gaussian broadened width is used.}
\end{figure}

\section{Conclusions}

In summary, the geometrical and magnetic properties of bimetallic Co nanowires coated with Cu have been studied by a genetic algorithm with a Gupta-like many-body potential and a spin polarized tight-binding Hamiltonian. The main conclusions can be made as follows. In the specific stoichiometric compositions, Cu atoms occupy the surface, while Co atoms prefer to occupy the interior of the nanowires to form the perfect-coated structures. The coated Cu atomic layer leads to large variations of the magnetic moment of Co nanowires, depending on the structure and the thickness of Cu layers. Single Co atom row in the center of nanowire is found to be nonmagnetic when coated with two Cu layers, while all the rest Co nanowires coated with Cu shell are still magnetic and their moments are
reduced.

\begin{acknowledgements}
This work is financially supported by the National Natural Science Foundation of China(No.29890210, 100230017). 
\end{acknowledgements}

\ \\
$^*$ E-mail: hyblwang@pub.hy.jsinfo.net \\
$^{\dagger}$ Current address: Institute for Shock Physics, Washington State University, Pullman, WA 99164. E-mail: jzhao@wsu.edu

\end{multicols}
\end{document}